# Space-Time Block Coded Spatial Modulation Aided mmWave MIMO with Hybrid Precoding

Taissir Y. Elganimi and Ali A. Elghariani
Electrical and Electronic Engineering Department, University of Tripoli
Tripoli, Libya
E-mail: {t.elganimi, a.elghariani}@uot.edu.ly

*Abstract*—In this paper, a combination of Space-Time Block Coded Spatial Modulation with Hybrid Analog-Digital Beamforming (STBC-SM-HBF) for Millimeter-wave (mmWave) communications is proposed in order to take advantage of the merits of Spatial Modulation (SM), Space-Time Block Codes (STBC), Analog Beamforming (ABF), and digital precoding techniques while avoiding their drawbacks. This proposed system benefits from the multiplexing gain of SM, from the transmit diversity gain of STBC, and from the Signal-to-Noise Ratio (SNR) gain of the beamformer. The simulation results demonstrate that the Zero Forcing (ZF) and the Minimum Mean Square Error (MMSE) precoded STBC-SM systems have better Bit Error Rate (BER) performance than the precoded SM systems. Moreover, the precoded SM shows a performance degradation compared to STBC-SM system. Furthermore, the BER is significantly improved by employing an array of ABF. In addition, it is demonstrated that a minimum of 2 antenna elements in the proposed scheme of STBC-SM-HBF are required to obtain better BER than that of the conventional SM and STBC-SM systems under the same spectral efficiency of 2 *bits/s/Hz*.

I. INTRODUCTION

IN wireless communications, Multiple-input Multiple-output (MIMO) transmission is an effective technique in improving capacity, reliability [1], and enhancing spectral efficiency of the next generation wireless systems [2]. However, with the availability of huge bandwidth in Millimeter-wave (mmWave) communication in the range from 30 GHz to 300 GHz, mmWave MIMO becomes an important candidate for the Fifth Generation (5G) wireless systems. mmWave communications was considered previously for outdoor point-to-point backhaul links [3], and for carrying indoor multimedia streams of high resolution [4].

Recently, there are still noticeable research work on the two general MIMO transmission strategies, namely; (1) the Space-Time Block Code[1] (STBC), which is proposed by Alamouti in [5] to exploit the potential of MIMO systems due to its low decoding complexity and implementation simplicity, and to achieve a full transmit diversity gain for two transmit antennas, and (2) the Spatial Multiplexing (SMX) which requires a transmit Radio Frequency (RF) chain for each transmit antenna. It is based on activating all transmit antennas for data transmission or reception [6].

On the other hand, Spatial Modulation (SM), which was first proposed by Mesleh *et al.* in [7], is based on activating only one transmit antenna at a time to reduce Inter-channel Interference (ICI) and the Inter Antenna Synchronization (IAS) between multiple antennas in MIMO systems. SM systems have the potential to reduce the energy consumption as well as the transmitter cost of MIMO systems. Antenna indices, at any instant, are used as means of data transmission in SM MIMO systems to convey the information bits.

In wireless communication systems, implementing the Analog Beamforming (ABF) with other techniques is an attractive way in order to achieve lower cost and power consumption. ABF can combat the limitation of the high propagation loss in mmWave communications, especially in an urban scenario [8]. On the other hand, Hybrid Analog-Digital Beamforming (HBF) is another promising technique which has been recently adopted in massive MIMO systems [9, 10]. It is based on combining the ABF with MIMO digital precoding technique, known as Digital Beamforming (DBF). HBF is a potential technique in mmWave 5G wireless systems to provide improved coverage, reduced RF cost and higher spectral efficiency.

*The main contribution of this paper is to implement and combine both ABF and linear precoding techniques together into HBF regime with SM and STBC-SM systems for the emerging mmWave communications. We refer to this system as STBC-SM-HBF. In order to underline the achieved gains, the BER performance of the proposed scheme is compared to the performance of the conventional SM technique. Computer simulations showed that the proposed STBC-SM-HBF scheme has better error performance over the classic SM with an optimal decoder due to its diversity advantage and the Signal-to-Noise Ratio (SNR) gain improvement achieved by the beamformer.*

To this end, the rest of this paper is organized as follows. In section II, the STBC-SM-HBF system model is introduced. Then, section III provides the simulation results of the BER performance, then the paper is concluded in section IV.

## II. SYSTEM MODEL

### A. STBC-SM Scheme

In Alamouti STBC, a complex symbol pair ($x_1$ and $x_2$) are taken from an $M$-ary Phase Shift Keying ($M$-PSK) or $M$-ary Quadrature Amplitude Modulation ($M$-QAM) constellation, where $M$ is the constellation size, and then transmitted from two transmit antennas in two symbol intervals orthogonally by the code word [5]:

$$\mathbf{X} = (\mathbf{x}_1 \quad \mathbf{x}_2) = \begin{pmatrix} x_1 & x_2 \\ -x_2^* & x_1^* \end{pmatrix} \quad (1)$$

where rows and columns correspond to the symbol intervals and the transmit antennas, respectively. For the STBC-SM scheme that proposed by Basar *et al.* in [11], the matrix in (1) is extended to the antenna domain. For example, STBC-SM with four transmit antennas and Binary Phase Shift Keying (BPSK) modulation scheme, it transmits the Alamouti STBC using one of the following four codewords [11]:

$$\begin{aligned}
\chi_1 &= \{\mathbf{X}_{11}, \mathbf{X}_{12}\} \\
&= \left\{ \begin{pmatrix} x_1 & x_2 & 0 & 0 \\ -x_2^* & x_1^* & 0 & 0 \end{pmatrix}, \begin{pmatrix} 0 & 0 & x_1 & x_2 \\ 0 & 0 & -x_2^* & x_1^* \end{pmatrix} \right\} \\
\chi_2 &= \{\mathbf{X}_{21}, \mathbf{X}_{22}\} \\
&= \left\{ \begin{pmatrix} 0 & x_1 & x_2 & 0 \\ 0 & -x_2^* & x_1^* & 0 \end{pmatrix}, \begin{pmatrix} x_2 & 0 & 0 & x_1 \\ x_1^* & 0 & 0 & -x_2^* \end{pmatrix} \right\} e^{j\theta}
\end{aligned} \quad (2)$$

where $\chi_i$, $i = 1,2$ are the STBC-SM codebooks, and each codebook contains two codewords $\mathbf{X}_{ij}$, $j = 1,2$ that do not interfere to each other. The resulting STBC-SM code is $\chi = \bigcup_{i=1}^{2} \chi_i$. And $\theta$ is the rotation angle that has to be optimized for a modulation scheme in order to ensure maximum diversity and coding gain due to the expansion of the signal constellation. An overlapping columns of the codeword pairs from different codebooks would occur, and hence reducing the transmit diversity order to unity if a rotation angle is not considered [11].

If four information bits $(u_1, u_2, u_3, u_4)$ are transmitted in two consecutive symbol intervals by STBC-SM system, the mapping rule for $2 \ bits/s/Hz$ transmission using BPSK modulation and four transmit antennas states that the first two information bits $(u_1, u_2)$ determine the antenna-pair position $l$, while the last two bits $(u_3, u_4)$ are used to determine the BPSK symbol pair. If a higher modulation order is used, for instance a modulation size of $M$, there will be four different codewords, each having $M^2$ different realizations. Therefore, the spectral efficiency of STBC-SM scheme for four transmit antennas is expressed as $m = (1/2) \log_2 4M^2 = 1 + \log_2 M \ bits/s/Hz$, where the factor $1/2$ normalizes for the two channel uses of the matrices in (2) [11].

In general, for a MIMO system with $N_T$ transmit antennas and $N_R$ receive antennas, the number of transmit antennas in STBC-SM scheme do not need to be an integer power of 2 as in SM scheme. The possible pairwise combinations are chosen from the $N_T$ transmit antennas for STBC transmission which provides a flexibility for STBC-SM system design. Therefore, the total number of STBC-SM codewords has to be an integer power of 2 and can be expressed as $c = \left\lfloor \binom{N_T}{2} \right\rfloor_{2p}$, where $p$ is a positive integer [11]. Moreover, the number of codewords in each STBC-SM codebook $\chi_i$, $i = 1,2,...,n-1$ can be calculated as $a = \lfloor N_T/2 \rfloor$, while the total number of codebooks is $n = \lceil c/a \rceil$. From $c$ antenna combinations, the resulting spectral efficiency of STBC-SM system can be calculated as [11]:

$$m = \frac{1}{2} \log_2 c + \log_2 M \quad (bits/s/Hz) \quad (3)$$

The general block diagram of the proposed STBC-SM-aided mmWave MIMO with hybrid precoding is depicted in Figure (1) where two transmit antennas are selected to transmit each Alamouti STBC symbol matrix. Thus, $2m = \log_2 c + 2 \log_2 M$ bits enter the STBC-SM transmitter during each two consecutive intervals. The first $\log_2 c$ bits determine the antenna-pair position $l = u_1 2^{\log_2 c - 1} + u_2 2^{\log_2 c - 2} + \cdots + u_{\log_2 c} 2^0$ that is mainly associated with the corresponding antenna pair. While the last $2 \log_2 M$ bits determine the STBC complex symbol pair ($x_1$ and $x_2$). In this scheme, each transmit antenna has $L$ array elements for the sake of achieving beamforming. The digital transmit precoder is employed in order to support the aforementioned digital beamforming, and/or to reduce the receiver's complexity.

### B. Zero-Forcing Precoding

In general, ZF linear precoders are widely used in MIMO systems with multiple transmit antennas. The main goal of using ZF precoders in MIMO systems is to remove both the inter-symbol interference (ISI) and the co-channel interference among the transmit antennas. However, it can be used to obtain the channel inverse at the transmitter when the number of transmit and receive antennas are equal. The general structure of ZF precoder with the knowledge of channel $H$ at the transmitter is written in terms of its Hermitian transposition and the inversion operation as follows [12]:

$$\mathcal{P}_{ZF} = H^H [HH^H]^{-1} \quad (4)$$

### C. MMSE Precoding

MMSE precoders can be obtained by applying the well-known MMSE receiver at the transmitter side. In this kind of precoders, the base station is assumed to be known in addition to the channel state information. Therefore, the linear MMSE precoder tries to find a good tradeoff between the interference and the noise, and it can be expressed as follows [13]:

$$\mathcal{P}_{MMSE} = H^H [HH^H + \sigma^2 I]^{-1} \quad (5)$$

where $\sigma^2$ is the noise variance, and $I$ is the identity matrix.

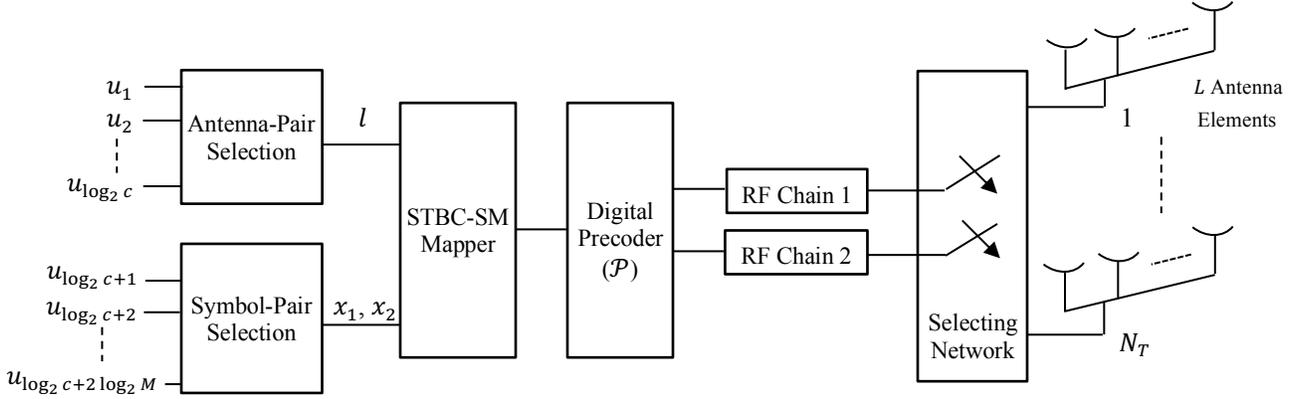

Fig 1. Block Diagram of the Proposed STBC-SM-aided mmWave MIMO Scheme with Hybrid Analog-Digital Beamforming.

## D. The Criteria of Generating ABF Weights

In this paper, the ABF is controlled based on the Angle of Departure (AoD), $\theta_{AoD}$, at the transmitter. This paper adopts a Uniform Linear Array (ULA) where the ABF weight, $\mathbf{w}_{T_i}$, with $L$ elements at each transmit antenna, is modeled as follows [9]:

$$\mathbf{w}_{T_i} = \begin{bmatrix} 1 & e^{-j\delta_T^{(i)}} & e^{-j2\delta_T^{(i)}} & \ldots & e^{-j(L-1)\delta_T^{(i)}} \end{bmatrix}^T \quad (6)$$

where $\delta_T^{(i)}$ is the electrical phase shift between each two antenna elements along the transmit antenna array that is expressed as $d.(2\pi/\lambda).\sin(\theta_{AoD}^{(i)})$. Also $\theta_{AoD}^{(i)}$ denotes the AoD towards the $i$th ABF of the transmitter. $d$ is the antenna spacing between each two antenna elements in each ABF, and $\lambda$ is the transmission wavelength.

This ABF weight is known as the ABF weight vectors with $L$ complex conjugate coefficients at the transmitter, and it contains the information about all antenna elements and the Direction of Arrival (DoA) of the transmitted signals.

## E. SM-HBF Receiver

Consider a $(N_T \times N_R)$-element MIMO system, where $N_T$ is the number of transmit antennas, and $N_R$ is the number of receive antennas, and let $H \in \mathbb{C}^{N_R \times N_T}$ denote the zero-mean and unity power channel matrix between the transmit and receive antennas, and $n \in \mathbb{C}^{N_R \times 1}$ is a zero-mean Additive White Gaussian Noise (AWGN) of power $N_O$.

In the general model of a SM system, a row vector $q(k)$ of $n_{SM} = (\log_2 N_T + \log_2 M)$ information bits is first transmitted, then it is mapped into another row vector $x(k)$ of size $N_T$ with It has all zeros except one symbol denoted by $x_l$. This symbol is transmitted over a MIMO channel $H(k)$ from the transmit antenna number $l$. The channel can be expressed as a set of column vectors $h_v$'s in a row vector $H$ as follows [7]:

$$H = [h_1\ h_2\ h_3\ \ldots\ h_{N_T}] = \begin{bmatrix} h_{1,1} & \cdots & h_{1,N_T} \\ \vdots & \ddots & \vdots \\ h_{N_R,1} & \cdots & h_{N_R,N_T} \end{bmatrix} \quad (7)$$

where each $h_v = [h_{1,v}\ h_{2,v}\ \ldots\ h_{N_R,v}]^T$ is a column vector represents the gain of the channel path between the transmit antenna $v$ and the receive antennas.

After employing the digital precoding, the received SM signal with activating a single antenna can be expressed as:

$$y(k) = H_{b\,(v=l)} x_l + n(k) \quad (8)$$

where $H_b = HP \in \mathbb{C}^{N_R \times N_R}$ is the multiplication of the channel matrix $H \in \mathbb{C}^{N_R \times N_T}$ to the precoding matrix $\mathcal{P} \in \mathbb{C}^{N_T \times N_R}$ and known as the effective channel.

The received signal after employing ABF at the transmitter of the precoded SM system can be expressed as:

$$y(k) = H_{P\,(v=l)} x_l + n(k) \quad (9)$$

where $H_P \in \mathbb{C}^{N_R \times N_R}$ represents the hybrid precoding matrix which is expressed as:

$$H_P = \sum_{i=0}^{L-1} \mathbf{w}_{T_i}^H H \mathbf{w}_{T_i} \mathcal{P} \quad (10)$$

In SM-HBF systems with the presence of the linear precoder $\mathcal{P}$, the received signal is normalized by taking into account the number of antenna elements with a factor of $1/L$.

In [14], A. Younis *et al.* have used Maximum Likelihood (ML) detectors for SM systems. In this detection algorithm, the received signal is used to achieve the prime goal of ML detectors which minimizes the Euclidean distance. Thus, estimating the transmit antenna number $\tilde{l}$ and the transmitted

symbols $\tilde{x}_l$ in the precoded SM system with $N_R$ receive antennas can be expressed as [14]:

$$[\tilde{l}, \tilde{x}_l] = \arg\min \left\{ \sum_{i=1}^{N_R} |y_i - H_{b_{l,i}} s|^2 \right\} \quad (11)$$

where $l \in \{1,2,3, \dots N_T\}$ and $s \in \{s_1, s_2, s_3 \dots s_M\}$ is all possible transmitted vectors, $H_{b_{l,i}}$ and $y_i$ are the $i$-th entry of $H_{b_l}$ and $i$-th entry of $y$ respectively.

### F. STBC-SM-HBF Receiver

Combining HBF with STBC-SM system can be obtained by using ABF with the precoded STBC-SM system considering that the channel is known at the transmitter side. In this paper, ZF and MMSE precoders are used with ABF to design STBC-SM-HBF scheme as modeled in Figure (1). Combining the linear precoding with ABF in STBC-SM system includes the weight vector and its Hermitian transpose operation. Therefore, the received sample vector of STBC-SM-HBF scheme can be expressed as:

$$y = \sum_{i=0}^{L-1} \sqrt{\frac{\rho}{\mu}} w_{T_i}{}^H \mathcal{H}_\chi w_{T_i} \mathcal{P} \begin{bmatrix} x_1 \\ x_2 \end{bmatrix} + n \quad (12)$$

In this equivalent channel model, $y$ represents the $2N_R \times 1$ equivalent received signal, and $n$ is the $2N_R \times 1$ noise vector which denotes the additive white Gaussian noise having the variance of $N_O$, while $\mathcal{H}_\chi$ denotes the $2N_R \times 2$ equivalent channel matrix of the Alamouti STBC-SM scheme that has $c$ different realizations according to STBC-SM codewords. In addition, $\mu$ is the normalization factor to ensure that $\rho$ is the average SNR at each receive antenna. Like SM-HBF systems, the received signal of STBC-SM-HBF is normalized by taking into account the number of antenna elements with a factor of $1/L$. Finally, the receiver estimates the transmitted symbols and the indices of the two transmit antennas that are used in the STBC transmission based on the ML criterion as presented in [11] with taking the effective channel matrix $H_b = \mathcal{H}_\chi P \in \mathbb{C}^{N_R \times N_R}$ into consideration.

### III. PERFORMANCE RESULTS AND COMPARISONS

In this section, the performance results of STBC-SM system for $2\ bits/s/Hz$ transmission is provided assuming Alamouti STBC transmission scheme of $4 \times 4$ MIMO system with ABF and HBF techniques. The BER performance of STBC-SM systems is compared to the conventional $2 \times 4$ SM systems. In addition, SM systems are compared to the Vertical Bell Labs Layered Space-Time Architecture (V-BLAST) which is the most basic form of MIMO detection algorithms. Throughout the simulation, the corresponding transmission wavelength was $\lambda = 0.5\ cm$ when ABF is employed where a carrier frequency of $60GHz$ is considered. It is assumed that the array elements are separated by half the wavelength $\lambda/2$, and a flat Rayleigh fading channel model is employed. Furthermore, the implementation of omnidirectional antenna elements is assumed to be employed at the transmitter.

### A. Comparison between SM, STBC-SM, V-BLAST, Precoded SM, and Precoded STBC-SM Schemes

Figure (2) presents the BER performance of $2 \times 4$ SM, $4 \times 4$ STBC-SM, $4 \times 4$ V-BLAST, and ZF and MMSE precoded SM and STBC-SM systems for $2\ bits/s/Hz$ transmission. ML detection and BPSK modulation are used in all of these systems. It can be observed that for the same spectral efficiency, STBC-SM outperforms V-BLAST and SM systems by about $4.5dB$, and $5dB$ respectively at the BER of $10^{-5}$ due to the diversity gain achieved by using Alamouti STBC in SM systems. On the other hand, $2 \times 4$ SM system shows a performance degradation of about $0.5dB$ at the BER of $10^{-5}$ as compared to $2 \times 4$ V-BLAST system. This finding shows that both systems perform almost the same with activating only one transmit antenna in SM system at a time, and hence reducing the cost significantly.

ZF and MMSE precoded SM showed a performance degradation of about $16dB$ as compared to the conventional SM system at the BER of $10^{-5}$. In contrast, the precoded SM system using MMSE precoder showed a little performance degradation compared to the ZF precoded SM system in the case of SNR lower than $13dB$. Furthermore, ZF and MMSE precoded STBC-SM systems showed a performance degradation of about $2dB$ and $4dB$ as compared to STBC-SM system at the BER of $10^{-5}$ respectively. This is due to the use of linear precoding strategies in SM systems that cause a wastage in the transmission power, and hence a worse BER performance is attained in these systems. In addition, the ZF precoded STBC-SM system showed a significant improvement of about $19dB$ at the BER of $10^{-5}$ as compared to the precoded SM systems which have the worst error performance in this comparison.

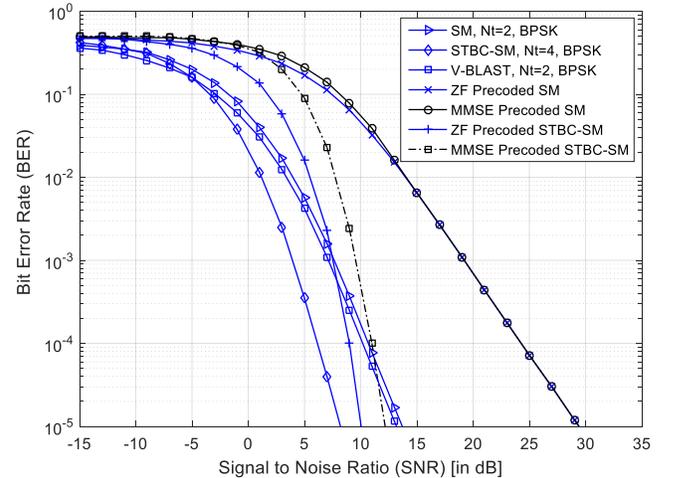

Fig 2. BER Performance of $2\ bits/s/Hz$ Transmission for SM, STBC-SM, V-BLAST, and Precoded SM and STBC-SM Schemes.

### B. Comparison between SM-ABF and STBC-SM-ABF Schemes

The BER performance versus SNR of SM and STBC-SM systems with ABF are compared in Figure (3). It shows the BER

performance improvement achieved by increasing the number of beam-steering elements of each antenna array at the transmitter of SM and STBC-SM systems. For SM-ABF systems, a performance improvement is obtained and SNR gains of approximately $6dB$, $9.5dB$, and $12dB$ are achieved by SM-ABF scheme as compared to the conventional SM system with single element at the BER of $10^{-5}$ for $L = 2, 3,$ and 4 elements, respectively. Similarly for STBC-SM-ABF systems with the same SNR gains over the conventional STBC-SM system. Thus, it can be said that STBC-SM-ABF scheme for $L = 2, 3,$ and 4, outperforms SM and SM-ABF systems by a transmit diversity gain of about $> 5dB$ at the BER of $10^{-5}$.

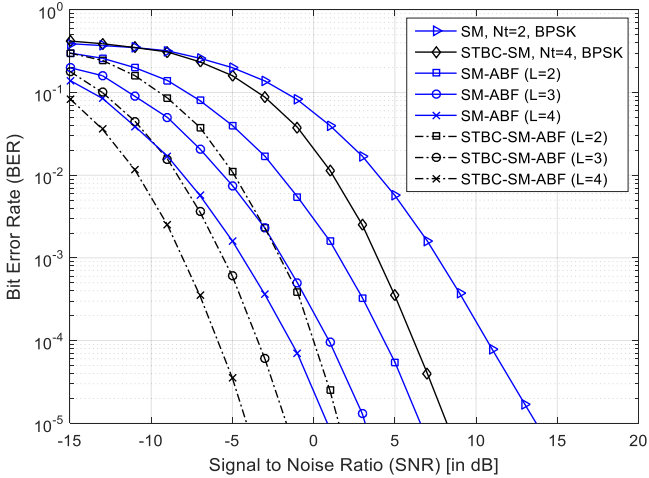

Fig 3. BER Performance of $2\ bits/s/Hz$ Transmission for SM and STBC-SM Schemes with Analog Beamforming.

*C. Comparison between SM and STBC-SM Systems with Digital and Hybrid Precoding*

As previously discussed, the precoded SM systems have the worst error performance compared to the other schemes in this study due to the use of linear precoding technique in SM systems. In Figure (4), a combination of SM and HBF systems is evaluated for a spectral efficiency of $2\ bits/s/Hz$. This figure shows that SM-HBF provides inferior BER performance to that of the conventional SM system. However, SM-HBF performance can be further improved by increasing the number of array elements, as depicted in the same figure. As a consequence, it is necessary to consider an improving technique to achieve larger SNR improvement than that provided by the SM-HBF scheme. Thus, combining STBC with SM system in the presence of HBF is proposed in this paper, and the attainable BER performance of this scheme is shown in Figure (5) for a spectral efficiency of $2\ bits/s/Hz$ for different values of beam-steering elements. It shows that by using STBC-SM systems with HBF, we can obtain better BER performance compared to that of SM-HBF and the conventional SM systems. For instance, the proposed scheme of STBC-SM-HBF provides almost $3.5dB$, $7.5dB$, and $10dB$ gains compared to the conventional STBC-SM systems at the BER of $10^{-5}$ with $L = 2, 3,$ and 4 elements, respectively. It is also obvious from these results that the proposed STBC-SM-HBF scheme achieves better BER performance than that of the conventional SM system even with only 2 antenna array elements. Thus, STBC-SM-HBF scheme offers an improvement in the BER performance compared to SM systems.

On the other hand, by comparing Figures (4) and (5), it is clear that SM-HBF schemes showed a performance degradation of about $9.5dB$, $6dB$ and $3.5dB$ as compared to the conventional SM system at the BER of $10^{-5}$ with $L = 2, 3,$ and 4 elements, respectively. The most striking feature of the proposed STBC-SM-HBF scheme is that it showed a performance improvement of almost $9.5dB$, $13dB$ and $15.5dB$ over the conventional SM system at the BER of $10^{-5}$ with $L = 2, 3,$ and 4 elements, respectively. This shows that the proposed scheme of STBC-SM-HBF has better error performance than the other schemes, and using STBC transmission in SM systems with HBF provides about $19dB$ gain at the BER of $10^{-5}$ as compared to SM-HBF systems.

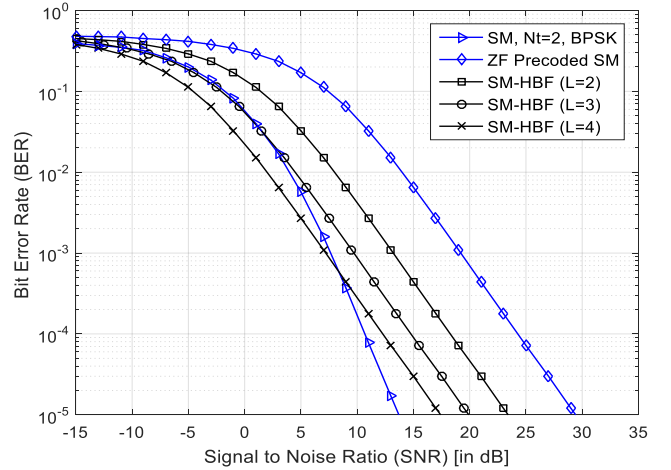

Fig 4. BER Performance of $2\ bits/s/Hz$ Transmission for SM Systems with Hybrid Analog-Digital Beamforming.

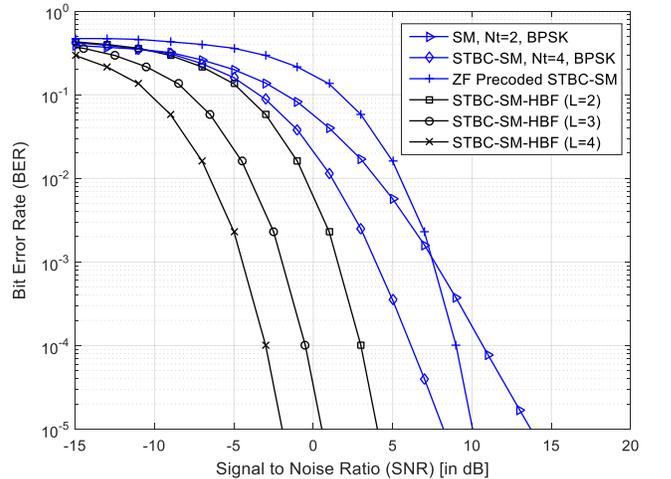

Fig 5. BER Performance of $2\ bits/s/Hz$ Transmission for STBC-SM Systems with Hybrid Analog-Digital Beamforming.

The SNR gain (in decibel) of ABF can be found approximately as:

$$20 \log_{10}(L/(L-1)) \quad (13)$$

This amount is the SNR improvement at a particular BER performance obtained by employing ABF with $L$ array elements at each transmit antenna in SM, STBC-SM, and ZF precoded systems as compared to the same systems with $L-1$ array elements at each transmit antenna.

*D. BER versus Number of Antenna Elements in SM-ABF, STBC-SM-ABF, SM-HBF, and STBC-SM-HBF Schemes*

In this subsection, the effect of BER as a function of the number of antenna elements for both SM and STBC-SM systems with ABF and HBF for $2\ bits/s/Hz$ transmission is studied at SNR of $-5dB$. It cab be observed from Figure (6) that STBC-SM-ABF system provides better error performance than the other schemes for different number of antenna array elements. While on the other hand, the BER performance of SM-HBF is the worst one among these systems. Therefore, employing antenna arrays in SM and STBC-SM systems yields better BER performance than the conventional SM and STBC-SM systems. Furthermore, SM-HBF and STBC-SM-HBF systems require more number of antenna elements than SM-ABF and STBC-SM-ABF systems respectively to achieve the same particular error performance, and hence higher transmitter cost due to the higher number of phase shifters required.

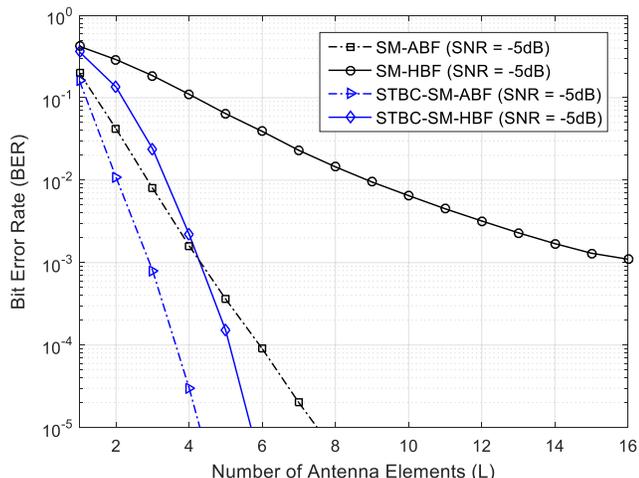

Fig 6. BER versus Number of Antenna Elements for $2\ bits/s/Hz$ Transmission in SM and STBC-SM Systems with Analog and Hybrid Beamforming.

Furthermore, the number of antenna array elements required in STBC-SM schemes to achieve a BER of $10^{-5}$ is less than that of SM schemes with ABF and HBF due to the benefits of implementing STBC in SM systems and the transmit diversity gain. Thus, less number of antenna elements is required in STBC-SM scheme for a particular BER performance, and hence less number of phase shifters at the transmitter. Therefore, this paper showed a significant improvement between the proposed scheme STBC-SM-HBF and SM-HBF due to combining STBC with SM systems.

IV. CONCLUSION

In this paper, STBC-SM-HBF scheme based on mmWave MIMO system is proposed. Results revealed that employing ABF provides SNR gain, and the achievable BER performance can be substantially improved as the number of beam-steering elements increases. Indeed, combining STBC with SM systems in the presence of hybrid precoding is a promising technique that showed a significant improvement in the error performance over SM and SM-HBF schemes. From the point of view of practical implementation, like the conventional SM scheme, the RF front-end of STBC-SM systems should be able to switch between different transmit antennas. Furthermore, only two RF chains are required in these schemes in which only two transmit antennas are employed to transmit information simultaneously. Unlike V-BLAST where all antennas are employed, and hence no need for synchronization in all of the transmit antennas in STBC-SM schemes. To conclude, STBC-SM with ABF and HBF schemes can be useful for low complexity, high-rate emerging wireless communication systems such as Long Term Evolution (LTE) and WiMAX systems, as well as for the future 5G mmWave communication systems.